\documentclass[preprint,aps,nofootinbib,showpacs,amsfonts,psfig]{revtex4}

\input{epsf.tex}

\usepackage{epsfig}

\newcommand{\be}{\begin{equation}}
\newcommand{\ee}{\end{equation}}
\newcommand{\bea}{\begin{eqnarray}}
\newcommand{\eea}{\end{eqnarray}}
\newcommand{\ba}{\begin{array}}
\newcommand{\ea}{\end{array}}

\begin{document}

\draft

\title{Lowered Nonextensive Stellar Distribution}

\author{J. M. Silva} \email{jmsilva@astro.iag.usp.br}
\author{R. E. de Souza} \email{ronaldo@astro.iag.usp.br}
\author {J. A. S. Lima} \email{limajas@astro.iag.usp.br}

\address{Departamento de Astronomia, Universidade de S\~ao Paulo, USP,
05508-900 S\~ao Paulo, SP, Brazil}

\begin{abstract}
The structure of globular clusters and elliptical galaxies are
described in an unified way through a new class of lowered models
inspired on the nonextensive kinetic theory. These power law models
are specified by a single parameter $q$ which quantifies to what
extent they depart from the class of lowered stellar distributions
discussed by Michie and King. For $q = 1$ the Michie-King profiles
are recovered. However, for $q<1$ there is a gradual modification in
the shape of the density profiles which depends on the degree of
tidal damage imposed on the model, thereby also providing a good fit
for globular clusters. It is also shown that a subclass of these
models, those with a deeper potential and $q$ slightly less than unity,
present a distribution resembling the de Vaucoulers $r^{1/4}$
profile which yields a good description of the structure of
elliptical galaxies. This subset of models follows this trend, with
a slight departure over nearly 10 orders of magnitudes.
\end{abstract}


\maketitle


\section{Introduction}

It is widely known that  the Maxwell-Boltzmann isothermal
distribution is somewhat truncated by the galactic tidal field
responsible for the capture of stars that become loosely bound to
the main structure. This means that the ideal distribution function
(DF) is depopulated due to the stars escaping from the system. As
shown by Michie \cite{MIC} to the case of non-rotating, spherically
symmetric systems, the resulting distribution may be obtained by
solving the Fokker-Planck equation. This class of distributions were
rediscussed by King \cite{Kin66} who showed that they give a good
representation for the observed density profiles of globular
clusters (GC). The phase space density for such models may be
expressed as \cite{Kin66,Kin78,Spitz87}

\begin{eqnarray}\label{e1}
f(\varepsilon) & = & {\rho_o}(2 \pi \sigma^{2})^{-3/2}
\left[e^{-\epsilon} - 1\right]
\;\;\;\;\;\;\;\;\;\;\;\;\;\;\;\;\;\;\;\;\; \varepsilon < 0\\
f(\varepsilon) & = & 0
\;\;\;\;\;\;\;\;\;\;\;\;\;\;\;\;\;\;\;\;\;\;\;\;\;\;\;\;\;\;\;\;\;\;\;\;
\;\;\;\;\;\;\;\;\;\;\;\;\;\;\;\;\;\; \varepsilon \ge 0 \nonumber
\end{eqnarray}
where the dimensionless quantity $\epsilon = ({1\over 2} v^2 +
\Phi({\bf r}))/{\sigma}^{2}$ is the energy of any given star in
units of the average thermal energy ($\Phi$ is the gravitational
potential and $\sigma$ is the velocity dispersion). Actually, in the
case of GC, some independent numerical analysis \cite{HM69,PK75,P76}
have confirmed the quantitative support provided by such
distributions usually called lowered models in the literature \cite{Binney}.

Unfortunately, these Michie-King models do not describe many
elliptical galaxies whose smoothness and symmetry are even more
striking than those presented by GC. Actually, instead of a specific
lowered distribution, the profiles of some giant elliptical galaxies
are very well described by the de Vaucouleurs $r^{1/4}$ law which
provides sometimes a remarkable fit over 10 orders of magnitudes of
surface brightness (see, for instance, de Vaucouleurs {\it et al.}
\cite{VC79} and references therein). Conversely, there are also some
elliptical galaxies (like NGC 4472) that are not fitted by the de
Vaucouleurs law as they are by the lowered Michie-King models. Such
a state of affairs is quite uncomfortable both from a methodological
and physical viewpoints. In particular, it opens a large window for
adjusting mechanisms whose primary objective is somewhat explain the
existence of individual profiles even for the spherically
symmetrical case.

On the other hand, an increasing  attention has been paid to
possible nonextensive effects in the fields of kinetic theory and
statistical mechanics \cite{Comp}(see http://tsallis.cat.cbpf.br/biblio.htm for
a regularly updated bibliography on this subject). The main
motivation is the lack of a comprehensive treatment including
gravitational and Coulombian fields, or more generally, any long
range interaction for which the assumed additivity of the entropy
present in the standard approach is not valid. Inspired on such
problems, Tsallis proposed the following $q$-parameterized
nonextensive entropic expression \cite{T88}

\begin{equation}\label {eq:Sq}
S_q = k_{B}\frac{\left[1-\sum_i p_i^{q}\right]}{(q-1)} \quad,
\end{equation}
where $k_B$ is the standard Boltzmann constant, $p_i$ is the
probability of the $i$-th microstate, and $q$ is a parameter
quantifying the degree of nonextensivity. This
expression has been introduced in order to extend the applicability of statistical mechanics
to system with long range interactions and has the standard Gibbs-Jaynes-Shannon entropy
as a particular limiting case (q = 1), namely
\begin{equation}\label{entrp.BG}
S=-k_{B}\sum_i p_i\ln p_i \quad.
\end{equation}

Ten years ago, the first attempts for exploring the kinetic route
associated to Tsallis entropy approach appeared in the literature.
The original kinetic derivation advanced by Maxwell \cite{Max1860}
was generalized to include power law distributions as required by
this enlarged framework \cite{SPL98} (see also \cite{BSL03}). In particular, it was
shown that the equilibrium  velocity $q$-distribution is given by
\begin{equation}
\label{eq0} f(v) = B_q \left[1-(1-q){m v^2\over 2 k_B
T}\right]^{1/(1-q)},
\end{equation}
where the $B_q$ is a $q$-dependent normalization constant which
reduces to the Maxwellian value for $q=1$. As shown in \cite{SPL98},
the above distribution is uniquely determined from two simple
requirements: (i) isotropy of the velocity space, and (ii) a
suitable nonextensive generalization of the Maxwell factorizability
condition, or equivalently, the assumption that $F(v)\neq
f(v_x)f(v_y)f(v_z)$. More recently, the kinetic foundations of the
above distribution were investigated in a deeper level through the
generalized Boltzmann's transport equation  which
incorporated the nonextensive effects using two different
ingredients. First, a new functional form to the kinetic local gas
entropy, and, second, a nonfactorizable distribution function for
the colliding pairs of particles whose physical meaning is quite
clear: the Boltzmann chaos molecular hypothesis is not valid in this
extended framework. It was also shown that the kinetic version of
the Tsallis entropy satisfies an $H_q$-theorem \cite{Lima01}, and 
still more important, the $q$-parameterized class of power law velocity
distributions emerged as the unique nonextensive solution describing
the equilibrium states. Several physical consequences
(in different branches) of the nonextensive kinetic theory have been
investigated in the literature, which includes its influence on the transport coefficients 
(spatially inhomogeneous dilute gas) of real gases and plasmas \cite{BSL03,Plasma}. In particular, a reasonable 
indication for a non-Maxwellian velocity power-law distribution from
plasma experiments have been demonstrated by Liu et al. \cite{Plasma1} and by Lima et al. \cite{Plasma2}. 
In high energy physics, many studies are illustrated by the solar neutrino problem \cite{HE1}, 
scattering processes in $e^{+}e^{-}$ annihilation \cite{Bediaga}, heavy  ion collisions \cite{alberico}, and the charm quark dynamics for
a thermalized quark-gluon plasma \cite{Walton}, among others (see \cite{Comp} for an extensive list of applications). 
Even experimental and theoretical deductions of the nonextensive q-parameter have been discussed  
in the literature by taking into account possible dynamical correlations or even the presence of 
inhomogeneities (due to clustering phenomena) thereby driving the system to a kind preequilibrium state \cite{HE4}. 
In this sense, the q-parameter cannot be thought as a fitting parameter, but a known quantity that should be ultimately determined by the correlations. 
The main lesson is that for systems with sufficiently complex dynamics (for instance, due to long range forces) 
other than Boltzmann-Gibbs statistics can  provide a better description.  

In the astrophysical context where our interest is  focused, Tsallis' statistics has been applied to
stellar collisionless systems \cite{PP93,DJ1}, peculiar velocity
function of galaxies clusters \cite{L98}, in studies of Jeans instability \cite{Lima02}, 
gravothermal instability \cite{TS2002} and the Chandrasekhar condition for the equilibrium and stability
of a  star \cite{DJ2}. There are also some expectations that such an approach may 
put some light on the universal structure of dark matter halos \cite{Hansen,DM,Feron}.

In a previous paper, we have
determined the radial and projected density profiles for two large
classes of isothermal stellar systems with basis on the equilibrium
power law q-distributions \cite{lima04}. Such models are based in
the following phase space density

\begin{equation}\label{e2}
f(\epsilon) = \frac{\rho_o C_{q}}{(2 \pi \sigma^{2})^{3/2}}\left[1 -
(1-q){\epsilon}\right]^{1/(1-q)}
\end{equation}

\noindent where the $q$-parameter quantifies the nonadditivity
property of the gas entropy. For generic values of $q \neq 1$, this
DF is a power law, whereas for $q=1$ it reduces to the standard
Maxwell-Boltzmann distribution function. Formally, this result
follows directly from the known identity $\rm{lim}_{d \rightarrow
0}(1 + dy)^{1 \over d} = {\rm{exp}(y)}$ as can be seen in
\cite{Abr72}. The quantity $C_q$ is a $q$-dependent normalization
constant given by \cite{Lima02}

\begin{equation}\label{e3}
C_q =
(1-q)^{1/2}\left(\frac{5-3q}{2}\right)\left(\frac{3-q}{2}\right)\frac{\Gamma
(\frac{1}{2}+{1\over 1-q})}{\Gamma({1\over
1-q})}\,\,\,\,\,\,\,\,\,\,\,\,
  q \leq 1
\end{equation}

\noindent and

\begin{equation} \label{e4}
C_{q}=(q-1)^{3/2}\left(\frac{3q-1}{2}\right)\frac{\Gamma\left(\frac{1}{1-
q}\right)}{\Gamma\left(\frac{1}{1-q}-\frac{3}{2}\right)}
\,\,\,\,\,\,\,\,\,\,\,\, \,\,\,\,\,\,\,\, q \geq 1,
\end{equation}

\noindent which reduce to the expected result in the limit $q=1$.

With basis on the ideas of Michie \cite{MIC} and King \cite{Kin66},
in this work we propose a new class of lowered nonextensive models
which are naturally associated with the above equilibrium
distribution. It will be explicitly assumed that the nonextensive
isothermal q-distribution discussed in \cite{Lima02} cannot strictly
be attained for a real stellar system due to the presence of tidal
effects with the mean local gravitational field, stellar encounters
or any other relaxation mechanism. This means that such mechanisms
must gradually modify the ideal velocity distribution in such a way
that the final distribution drop to zero at a finite velocity.
Hence, one may suppose that the stellar system approaches as far as
it can to a quasi power law final state which collectively can be
termed nonextensive lowered spheres. As we shall see, this new class
of models may potentially describe both classes of spherically
symmetric systems, namely globular clusters and elliptical galaxies,
thereby reinforcing the possibility that nonextensive effects may
have a considerable importance in the astrophysical domain.

The paper is organized as follows. Next section we set up the basic
equations defining the lowered nonextensive models. In section III
we discuss a large class of density profiles obtained trough a
numerical solution of the Poisson equation. Although reproducing the
Michie-King models exactly for $q\rightarrow 1$, we show that for
$q$ smaller than unity the standard profiles are moderately
modified, and therefore, they also provide a good representation for
the structure of GC. In section IV we discuss a specific q-lowered
model characterized by a relatively deep central potential. It
resembles the de Vaucouleurs law being therefore potentially useful
to describe the structure of elliptical galaxies. This model has an
extra bonus: it predicts the existence of small fluctuation similar
as that ones appearing in the profiles of early type galaxies, as
discussed by Caon \cite{Cao97} based on the S\'ersic
$r^{1/n}$ law. Finally, a summary and the main conclusions are
presented in section V.

\section{Lowered Stellar Distribution: Nonextensive Approach }

Let us now consider a class of lowered nonextensive q-distribution
defined by

\begin{eqnarray}\label{e5}
f(\varepsilon) & = & \frac{\rho_o C_{q}}{(2 \pi \sigma^{2})^{3/2}}
 \left[\left(1-(1-q) \varepsilon \right)^{1/(1-q)}-1\right]
 \;\;\;\varepsilon < 0\\
 f(\varepsilon) & = & 0
 \;\;\;\;\;\;\;\;\;\;\;\;\;\;\;\;\;\;\;\;\;\;\;\;\;\;\;\;\;\;\;\;\;\;\;\;
 \;\;\;\;\;\;\;\;\;\;\;\;\;\;\;\;\;\;\;\;\;\;\;\;
 \varepsilon \ge 0 \nonumber
\end{eqnarray}
which arises naturally from our earlier adoption of the isothermal
power law distribution to the unperturbed collisionless system (cf.
equations (1) and (2)). This means that to the accuracy of the
Michie-King approximation, the above distribution corresponds to the simplest
non-extensive extension representing the steady state solution of
the Fokker-Planck equation \cite{LSSN} (see also \cite{Tam1}). The corresponding mass density profile
can be obtained by integrating this distribution over the whole
possible energy interval. It thus follows that the density profile
is given by the expression

\begin{equation}\label{e8}
\rho= \frac{2{\rho_o} C_{q}}{\sqrt \pi}\int_{\phi}^0
\left[\left(1-(1-q) \varepsilon \right)^{\frac{1}{1-q}}-1\right]
(\varepsilon - \phi)^{1/2}d\varepsilon
\end{equation}
where $\phi = \Phi/\sigma^{2}$ is the dimensionless potential, and
the integration limits have been defined by keeping in mind that
only bounded objects are present in the stellar distribution. Once
again, this expression asymptotically approaches the lowered
Michie-King models when $q\rightarrow 1$. This expression linking
the mass density to the gravitational potential does not have a
closed analytical form and therefore must be numerically integrated.
Now, in order to simplify the numerical algorithm, it proves
convenient to introduce the variable, $\varepsilon = \chi \phi$,
with (\ref{e8}) assuming the form

\begin{eqnarray}\label{e9}
\rho = \frac{2{\rho_o} C_{q}}{\sqrt \pi}(-\phi)^{3/2}
\int_0^1\left[\left(1-(1-q)\phi\chi\right)^{\frac{1}{1-q}}-1\right]
(1-\chi)^{\frac{1}{2}}d\chi \;\;\;
\end{eqnarray}
which is more amenable to be solved by numerical discretization.
Another benefit from this expression is that we can easily verify
that the relation between the mass density and the gravitational
potential has two asymptotic regimes. In the first limit, when
$\phi\simeq 0$, we are close to the external surface and we can
expand the power expression appearing inside the brackets using the
binomial expansion to obtain

\begin{equation}\label{e10}
\rho \simeq \frac{8{\rho_o} C_{q}}{15\sqrt \pi}(-\phi)^{5/2}
\;\;\;\;\;\;\;\;\;\;\;\;\;\;\;\;\;\;\;\;\;\;\;\;\;\;\;\;\ \phi\simeq
0
\end{equation}

\noindent showing that the external structure of these models
correspond to a polytropic sphere with Lane-Emden index $n=5/2$. In
fact for $-\phi \ge 1$ this expression gives a good representation
of the model as we can verify from figure \ref{fig1}. In the
internal structure we can find another regime in those regions where
the gravitational potential is sufficiently large so that $(1-q)
\phi\chi >> 1$ in which case,

\begin{equation}\label{e11}
\rho \propto (-\phi)^{\frac{5-3q}{2(1-q)}}
\end{equation}

\noindent showing that in this limit the internal structure is also
described by a polytropic sphere  whose Lane-Emden index,
$n=(5-3q)/2(1-q)$, depends on the nonextensive parameter. In fact as
showed in \cite{lima04}, the nonextensive isothermal distributions
have a profile density corresponding exactly to this polytropic
structure. Therefore, in those models were the central potential is
sufficiently deep, the internal structure is closely described by
this approximation. In particular, we can verify that the model
$q=5/7$ corresponds to a polytropic sphere with $n=5$ which is the
limiting case dividing the Lane-Emden family in a branch having
finite mass and radius from those having infinite radius and mass.
Therefore, in the absence of tidal truncation, models satisfying the
restriction $5/7 <q\le 1$ have infinite mass and infinite radius and
are the ones that we will give more attention bellow. We can also
see that when $q=0$ the whole structure is exactly represented by  a
$n=5/2$ polytropic sphere. In figure \ref{fig1} we show dependence
of the mass density as a function of the gravitational potential for
a set of representative values of the $q$ index. We can see from
this plot that in the general case the whole structure can be
described as the result of a smooth transition between $n=5/2$ at
the external region to an internal region having $n=(5-3q)/2(1-q)$.
As happens in the Michie-King models, the value of the central
potential is also arbitrarily set to define implicitly the total
mass and the external radius. Therefore, only objects with a
sufficiently deep gravitational potential have the ability to show
the very internal limiting polytropic structure. In all the other
cases the internal region falls in the transition regime where the
external polytropic is gently changing towards the asymptotic
internal limit. In certain sense, this characteristic is responsible
by the variety of density profiles found in these models.

\section{The Integrated Density Profiles}

Having obtained the density as a function of the gravitational
potential we can proceed by solving the Poisson equation which
provides our dynamical link to obtain the density profile. Following
King \cite{Kin66} we normalize the density to its central value
$\rho_0$ and introduce a core radius defined as

\begin{equation}\label{e12}
r_c=\frac{9 \sigma^2}{4\pi G \rho_0},
\end{equation}

\noindent that will be used as a unit to measure the radial
distance. In the normalized radial coordinate, the Poisson equation
becomes

\begin{equation}\label{e13}
\frac{1}{r^2}\frac{d}{dr}r^2 \frac{d \phi}{dr}= 9 \varrho
\end{equation}

\noindent where the dimensionless density
$\varrho=\rho(\phi)/\rho(\phi_0)$ can directly be obtained from the
solution of equation (\ref{e9}) for each choice of the parameter
$\phi_0$, whose value will fix the whole structure of each model.

\begin{figure}[ptbh]
\centerline{\epsfysize=90mm\epsffile{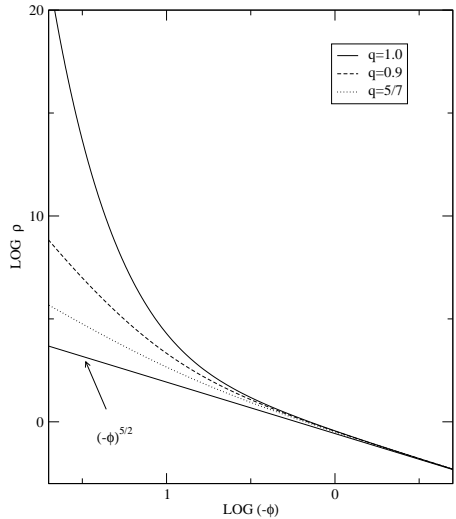}} \caption{The density
profiles as a function of the gravitational potential has two
regimes. In the external region the structure is approximately
described by a $n=5/2$ polytropic index for all models. In the
central region of those objects having a deep potential, the
limiting  structure depends explicitly on the nonextensive
parameter, and resembles a polytropic with
$n=(5-3q)/2(1-q)$.}\label{fig1}
\end{figure}

\begin{figure}[ptbh]
\centerline{\epsfysize=90mm\epsffile{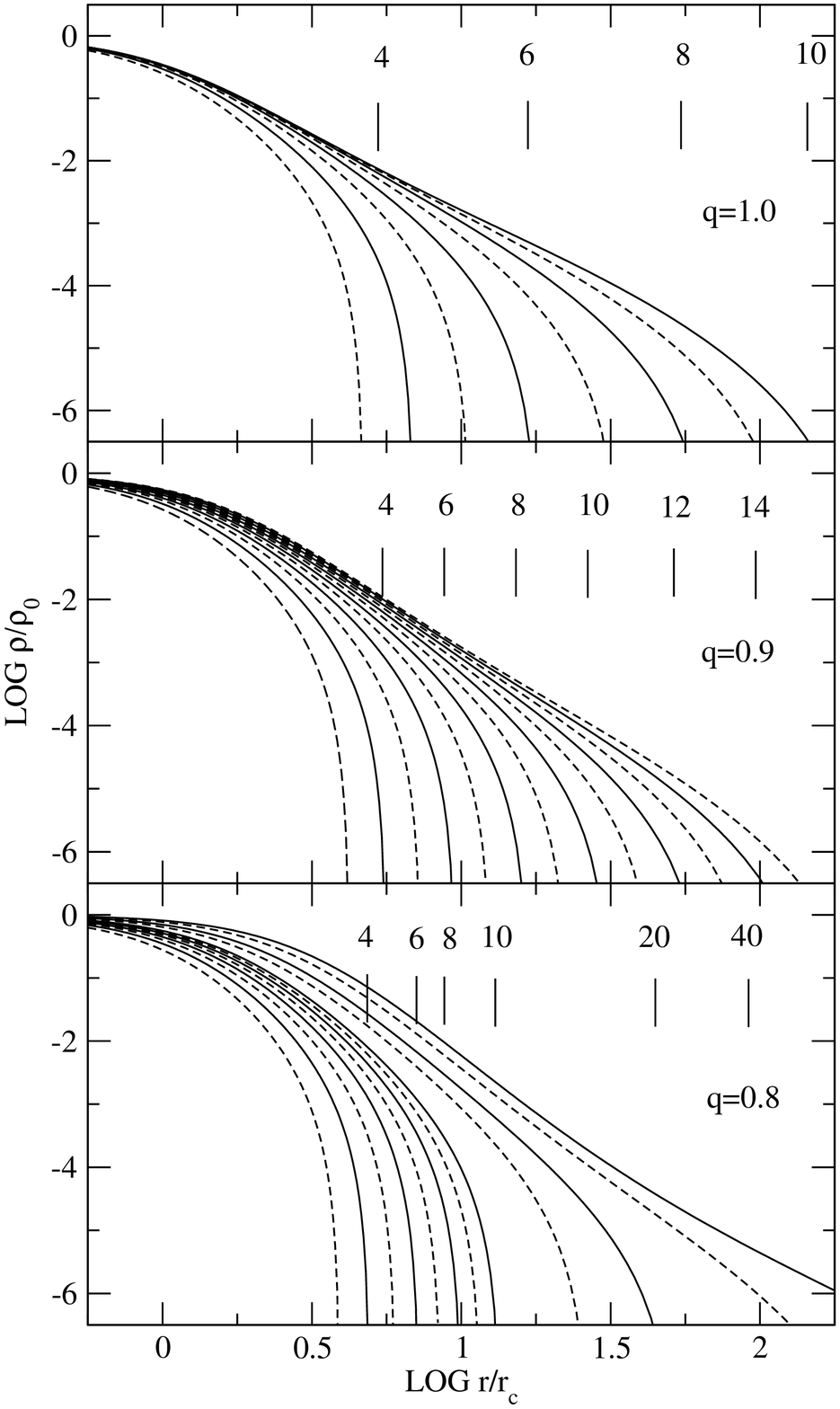}}\caption{Mass density
profiles for three representative models. The upper panel correspond
to a $q=1$ model which is exactly the standard Michie-King models.
The other two panels correspond to models with $q=0.9$ and $0.8$
respectively. We can see that models having the same size in units
of $r_c$, but different values of $q$, show a large variety of
density profiles. In each panel it is also indicated the
corresponding $-\phi_0$ of each model.} \label{fig2}
\end{figure}

\begin{figure}[ptbh]
\centerline{\epsfysize=100mm\epsffile{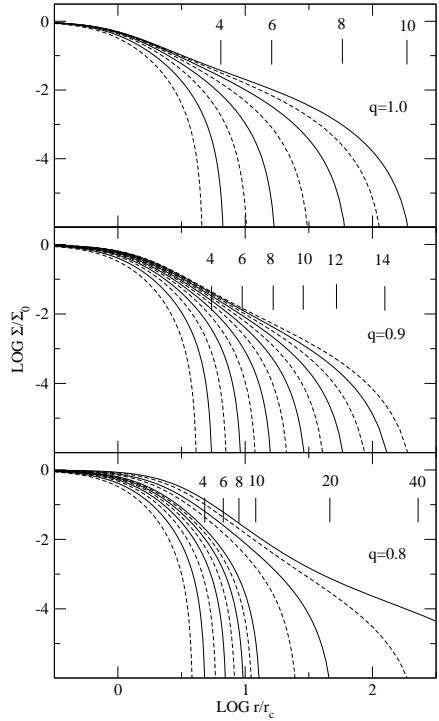}}\caption{ Projected
density profiles for the 3 classes of models presented in Figure
\ref{fig2}. Models corresponding to lower values of $-\phi_0$, and
hence $C$, have a higher similarity due to the dominant effect of
the external polytropic  structure ($n=5/2$).}\label{fig3}
\label{Fig3}%
\end{figure}

The Poisson equation can numerically be solved by imposing the
boundary conditions $\phi=\phi_0$ and $d\phi/dr=0$ at the origin
$r=0$. The solution proceed from the center and stop at the external
surface where $r=r_t$ and $\phi=\varrho=0$. The external and the
core radii define naturally a concentration index $C=r_t/r_c$
indicating how important is the truncation effect due to tidal
field. Models with low values of $C$ correspond to structures where
the tidal field had imposed a severe damage to the unperturbed
structure. In figure \ref{fig2} we present a set of solutions for
three values of $q$. In the upper panel we show the $q=1$ model,
corresponding to the usual class of lowered Michie-King models. In
the other two panels we present the profiles for $q=0.9$ and
$q=0.8$, showing that models with the same extent of the standard
lowered models have different density profiles. Note that all curves
have zero gradient for sufficiently small values of $r$ regardless
of the value of $q$. Indeed, such a behavior is more noticeable for
smaller values of $q$. In figure \ref{fig3} we present the projected
density profile for each model presented in figure \ref{fig2}.
Assuming that light traces mass these profiles should directly be
compared with the surface brightness profiles.

The total mass of the q-lowered models can be evaluated by
integrating the density profile as

\begin{equation}\label{e14}
M= 4 \pi r_c^3 \rho_o \int_0^C \varrho r^2 dr=4 \pi r_c^3 \rho_o
\mu,
\end{equation}

\noindent where $\mu$ is a dimensionless mass indicator which can be
obtained trough numerical integration. From Tables I and II we can
see that models with lower concentration $C$ representing severely
damaged structures tend to show a mass distribution close to an
uniform spheres with $\mu = 1/3$. On the contrary, in objects with
higher concentration index the mass is heavily dependent on the
central core distribution. In the course of the numerical solution
the integration begins by defining an arbitrary value for the
central potential $\phi_0$ and carrying out the integration to a
null value at the external surface. However, since we have estimated
the total mass we may use its value for determining the
gravitational potential at the external surface $-GM/r_t$ which can
be expressed in our units as

\begin{equation}\label{e15}
\phi_t=\Phi(r_t)/\sigma^2= -9 \mu/C
\end{equation}

\noindent and this value can be used as a correction to be added in
order to find the gravitational potential at the center. As a final
step, the corrected potential can be used to estimate the potential
energy

\begin{equation}\label{e16}
U = \frac{1}{2}\int_0^{r_t} 4\pi \Phi \rho r^2 dr =\frac{GM}{r_t}
\nu
\end{equation}

\noindent where

\begin{equation}\label{e17}
\nu = \frac{C}{18 \mu^2}\int_0^C \phi r^2 dr .
\end{equation}

\begin{table*}
\centerline{
\begin{tabular}{cccccccc}
$-\phi_0$& $r_h/r_c$& $\mu$& $\phi_t$& Log C& $\nu$& C$_1$& C$_2$\\
\hline \hline
     &     &      &      &q=1.0&     &     &     \\\\
-2   &0.567& 0.224&-0.631&0.505&1.399&1.609&1.486\\
-3   &0.651& 0.413&-0.790&0.672&1.565&1.641&1.527\\
-4   &0.701& 0.645&-0.840&0.840&1.803&1.688&1.587\\
-5   &0.729& 0.941&-0.791&1.030&2.178&1.760&1.678\\
-6   &0.744& 1.346&-0.673&1.255&2.733&1.880&1.821\\
-7   &0.751& 1.985&-0.529&1.528&3.489&2.141&2.036\\
-8   &0.755& 3.178&-0.419&1.834&4.166&2.512&2.241\\
-9   &0.757& 5.569&-0.381&2.119&4.176&2.883&2.217\\
-10  &0.757&10.019&-0.403&2.350&3.702&2.772&2.064\\
\hline
     &     &      &      &q=0.9&     &     &     \\\\
-2   &0.579& 0.224&-0.662&0.484&1.360&1.600&1.475\\
-3   &0.679& 0.410&-0.875&0.625&1.472&1.620&1.501\\
-4   &0.751& 0.629&-1.012&0.748&1.612&1.643&1.533\\
-5   &0.806& 0.880&-1.083&0.864&1.786&1.670&1.570\\
-6   &0.850& 1.165&-1.096&0.981&2.006&1.701&1.615\\
-7   &0.888& 1.487&-1.063&1.100&2.283&1.738&1.670\\
-8   &0.920& 1.852&-0.993&1.225&2.638&1.783&1.738\\
-9   &0.950& 2.274&-0.897&1.358&3.095&1.837&1.825\\
-10  &0.978& 2.771&-0.785&1.502&3.685&1.908&1.938\\
-11  &1.003& 3.375&-0.667&1.658&4.448&2.005&2.093\\
-12  &1.029& 4.142&-0.552&1.830&5.408&2.148&2.309\\
-13  &1.053& 5.171&-0.451&2.014&4.678&2.378&2.599\\
-14  &1.076& 6.625&-0.373&2.204&7.542&2.774&2.909\\
\hline
\end{tabular}}
\caption[]{}\label{tab1}
\end{table*}

\begin{table*}
\centerline{
\begin{tabular}{cccccccc}
$-\phi_0$& $r_h/r_c$& $\mu$& $\phi_t$& Log C& $\nu$& C$_1$& C$_2$\\
\hline \hline
     &     &      &      &q=0.8&     &     &     \\\\
-2   &0.588& 0.225&-0.688&0.468&1.329&1.594&1.467\\
-3   &0.699& 0.409&-0.942&0.592&1.407&1.607&1.484\\
-4   &0.784& 0.626&-1.144&0.692&1.494&1.620&1.503\\
-5   &0.855& 0.869&-1.300&0.779&1.589&1.633&1.522\\
-6   &0.915& 1.137&-1.418&0.858&1.695&1.647&1.542\\
-7   &0.969& 1.427&-1.503&0.932&1.811&1.660&1.562\\
-8   &1.017& 1.737&-1.560&1.001&1.939&1.673&1.583\\
-9   &1.062& 2.067&-1.595&1.067&2.078&1.686&1.604\\
-10  &1.104& 2.417&-1.609&1.131&2.231&1.698&1.626\\
-15  &1.286& 4.438&-1.490&1.428&3.247&1.760&1.742\\
-20  &1.442& 6.934&-1.193&1.719&4.949&1.820&1.881\\
-30  &1.708&14.190&-0.471&2.434&15.604&2.023&2.562\\
-40  &1.936&50.098&-0.139&3.510&31.581&17.164&3.918\\
\end{tabular}}
\caption[]{Resume of the lowered model parameters. The upper panel,
$q=1$, shows the same solution as founded with the King models. The
other panels illustrate the effect of changing the $q$ non
extensivity parameter on the structural parameter.}\label{tab1}
\end{table*}

In Tables I and II we show the major parameters obtained in this
class of lowered models. The first column present the uncorrected
central potential, $\phi_0$, used to start the integration of the
model. The radius containing half the mass, $r_h$, is shown in
column 2 in units of the core radius. For the King model, $q=1$, the
half radius is relatively stable even considering the large
variation covered by the concentration index. We see that the tidal
radius, $r_t$, varies by more than two orders of magnitudes while
the half radius is kept almost fixed at 70\% of the core radius. For
the other models this stability of the half radius is not generally
preserved. As an example, for a model with $q=0.9$ a similar
variation of the concentration index would imply in a variation of a
factor of two on the half radius. Actually, models with lower $q$
index show a more important departure from the Maxwellian case since
the variation of the half radius is even larger. This large
variation in the half radius is a consequence of a change in the
shape of the density profile for these models as can be seen from
figures \ref{fig1} and \ref{fig2}. In fact, these figures show that
for a similar external radius the gradient in the density profile is
different for each value of $q$.

In the last two columns of Tables I and II we present two
concentration indices based only on the projected surface density
distribution. They are based in the radii containing respectively
1/4, 1/2 and 3/4 of the total mass in each model. The concentration
index $C_1$ is defined to be equal to $r_{1/2}/r_{1/4}$ and
therefore measure the degree of concentration in the central mass
distribution. On the other hand the index $C_2$ is estimated as
$r_{3/4}/r_{1/2}$ and measure the spread of the mass distribution of
the outer region. These two indices are useful to give a broad
representation of the gamma of variations seen in the surface
density profiles for these models. For a single parameter family
each profile is represented as one single point in the $C_1$,$C_2$
plane. This feature is illustrated in figure \ref{fig4} for each
model represented in Tables I and II. For a given value of $q$ the
models falls in a line representing the gross features of the
individual profiles. Along each line $q$ line the position of the
points are uniquely determined by the parameter $-\phi_0$. On the
other hand these same plots can be easily obtained from luminosity
observed profiles. Therefore we can use the points in this plane to
give a gross diagnostic of the models more appropriated to each
class of objects. Observe that all models tend to converge to a
single point for low $-\phi_0$ models, independent of the $q$ index.
This is a consequence of the density profile as a function of the
potential as shown in figure \ref{fig1}. For low values of $-\phi_0$
the profiles tend to approximate a polytropic with $n=5/2$. As a
consequence, these models tend to present a unique surface density
profile and therefore also have a unique pair of concentration
indices $C_1\simeq 1.46$ and $C_2\simeq 1.28$. This regime coincides
with the profiles of highly tidally affected models where the tidal
radius is closer to the core radius. Therefore, strongly tidally
truncated GC tend to be equally well represented by different $q$
models. However, when we begin to sample larger values of the
central potential we observe the differences among the  $q$ models.
In this regime the individual density profiles behaves quite
differently as a function of the parameter $q$.

The King models, represented here by the subclass with $q=1$, are
known to give a good representation of the globular clusters. An
empirical fit to the radial distribution of GC show that
\cite{Djo94}

\begin{equation}\label{e19}
\rho(r) = \rho_0 ( 1+ \frac{r}{r_c})^{-\alpha}
\end{equation}

\noindent with $3.5< \alpha < 4$.  Using this empirical description
we indicate in figure \ref{fig4} the region corresponding to the GC
profiles by the letter G. We can see that models having $\alpha=3.5$
tend to fall quite close to the region described by the King models.
At this point one may ask if these q-lowered models may provide at
least a reasonable description for elliptical galaxies. Such a
possibility will be examined next section.

\section{Profiles For Elliptical Galaxies}

The density profile of elliptical galaxies have been widely
investigated either from the observational and theoretical point of
view. A very successful approach consist in represent the luminosity
profile of these objects by $r^{1/4}$, usually termed de Vaucouleurs
law \cite{V48,VC79}. Although existing many indications that such
empirical relation is not universal, there is a firm believe that it
provides a good representation for several elliptical galaxies.

A generalization of the de Vaucouleurs law is the so-called S\'ersic
law where the luminosity profiles are represented by an expression,

\begin{equation}
I_r =I_e 10^{-b_n[(\frac{r}{r_e})^{1/n}-1]}
\end{equation}

\noindent where $n$ is the S\'ersic index and $b_n$ is a properly
defined constant so that the effective radius $r_e$ contains half of
the total luminosity. In particular, for $n = 4$ the de Vaucouleurs
profile is recovered. Models with $n\simeq 3$ can be represented by
the lowered King models, but those with $n\simeq 4$ are definitely
better represented by a model with $0.9 < q < 1$.

\begin{figure}[ptbh]
\centerline{\epsfysize=100mm\epsffile{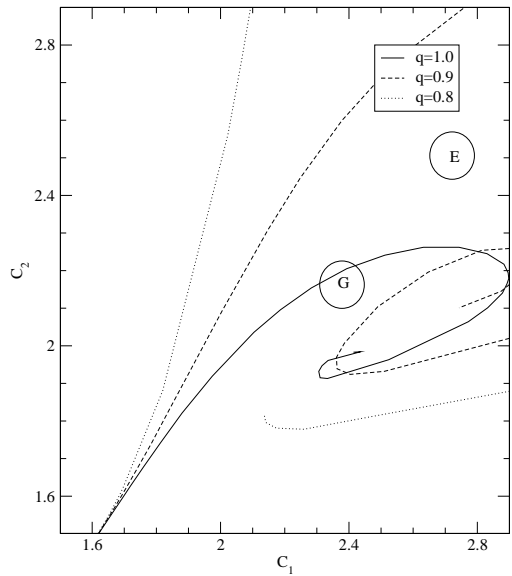}}\caption{This
concentration diagram shows the gross characteristic of the mass
distribution as a response in changing the $q$ nonextensive
parameter. The interrupted line with the label G mark the
approximated location of GC lying close to the King model,
corresponding to $q=1$. The circular region with the label $r^{1/4}$
law marks the position of the elliptical galaxies obeying the de
Vaucouleurs law. In this case, a good description of the luminosity
profile is provided by a model with $q=0.95$.} \label{fig4}
\end{figure}

\begin{figure}[ptbh]
\centerline{\epsfysize=100mm\epsffile{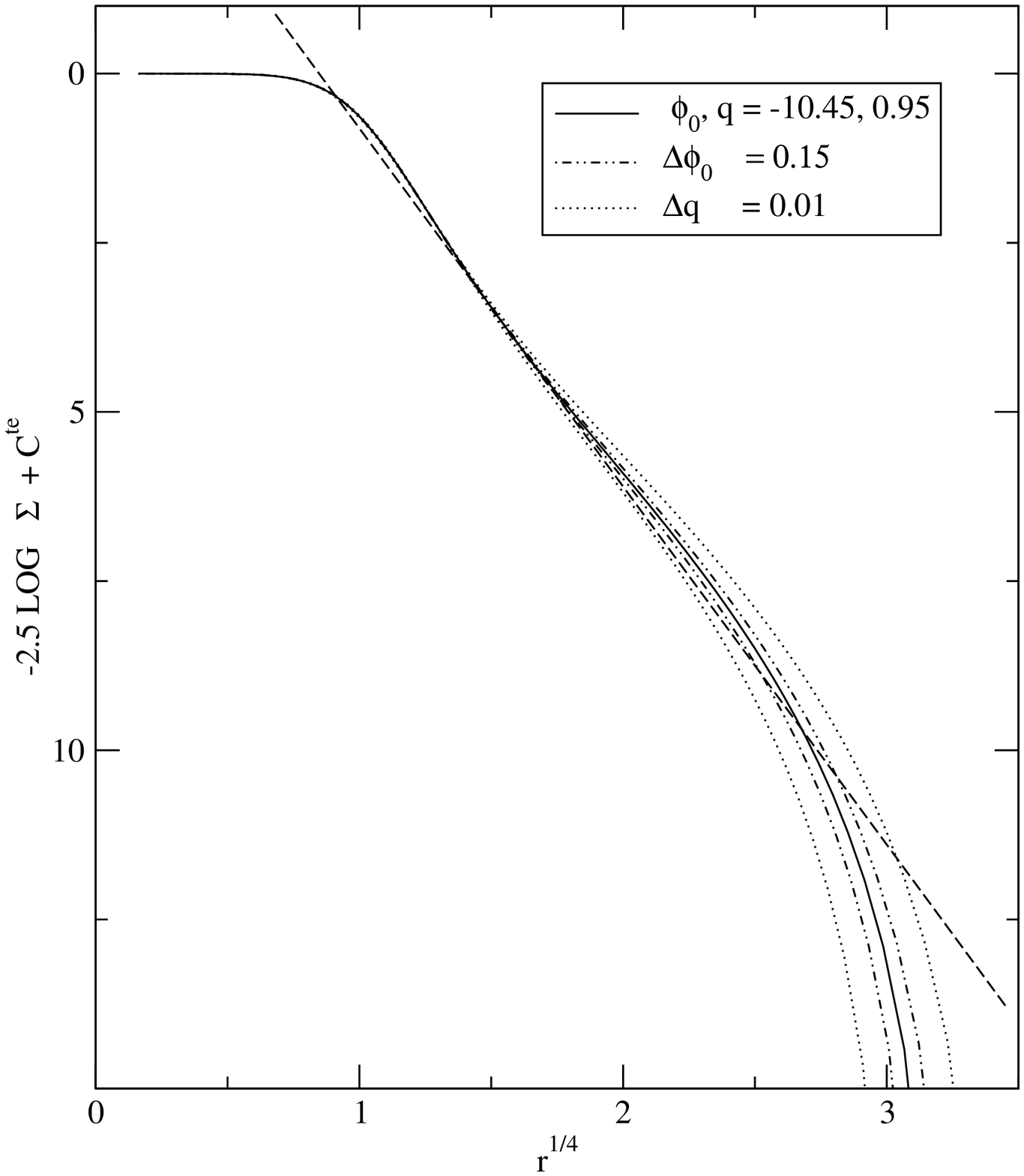}} \caption{ A
comparison of the $q=0.95$ model with the de Vaucouleurs law showing
that this model may give a good description of the $r^{1/4}$ law
over an interval of 10 magnitudes.} \label{fig5}
\end{figure}

A simple integration of the de Vaucouleurs law leads to profile
corresponding to a pair of concentrations indexes $C_1=2.730$ and
$C_2=2.506$, which is represented by the label $r^{1/4}$ in figure
\ref{fig4}. It is interesting to observe that this point is closely
represented by a model $q\simeq 0.95$ and $-\phi_0 \simeq -10.45$.
in figure \ref{fig5} we present the projected density profile
corresponding to this model. As one may see, this model is a good
representation of the de Vaucouleurs law over an interval of nearly
10 orders of magnitudes. There are two points that deserve a further
comment. The first one is that models with parameters $q = \simeq
0.95$ and $\phi=-10.45$ are the best representation of the de
Vaucouleurs law. If we explore the parameter region neighboring this
point we can verify that the agreement is progressively lost. In
figure \ref{fig4} we illustrate that point by exploring points
departing 1\% from that optimum model.

The other point of interest is that our representation of the de
Vaucouleurs law does not exactly reproduce the $r^{1/4}$ profile.
There is a noticeable oscillation around that trend. Over the 10
magnitude interval were the $q$ profile is close to the de
Vaucouleurs law this oscillations have an amplitude of the order of
0.2 magnitudes. The remarkable fact is that the same phenomenon was
observed in \cite{Cao97} in the elliptical galaxies belonging to the
Virgo cluster. This is clearly a point that deserve a closer
scrutiny  in the future.

\section{Conclusion}

In this paper we have proposed a new family of lowered models with
basis on the nonextensive kinetic theory. As we have seen, these
q-lowered models based on Tsallis statistics extend naturally the
class of Michie-King family and are also able to reproduce the
observed structure of globular clusters. Although giving a broader
range of density profiles than the usual King-Michie lowered models,
the region occupied by these system in the concentration index
diagram remains close to the $q=1$ models which correspond exactly
to the classical King models. It remains a matter of investigation
to verify if the actual observed profiles of galactic globular
clusters do favor a $q=1$ lowered Maxwellian distribution or if
there exist room do discuss the presence of departures from this
solution.

For elliptical galaxies the situation is much more favorable to the
nonextensive models. It is known that the Michie-King models do not
give a close representation of the observed profiles of elliptical
galaxies. Some objects, like NGC 4472, possibly affected by tidal
truncation are well represented by the King profile \cite{Kin78}.
However, the vast majority of bright elliptical, as NGC 3379, do no
fit quite well in the King model and tend to be closer to the de
Vaucouleuers profile \cite{VC79}. In this concern, the contribution
of the lowered nonextensive models can be quite important since it
is possible to find models which are closely resembling the
structure of elliptical galaxies. In fact a model with $q=0.95$ m is
able to reproduce the de Vaucouleurs empirical law over more than 10
magnitudes of surface brightness. It could be a matter of great
interest to pursue a more detailed comparison of these models with
the actual data on elliptical galaxies. In particular these non
extensive models predict a systematic departure from the de
Vaucouleurs law in the central region similar to the findings by
Ferrarese et. al \cite{Fe94}. Using data from the Hubble space
telescope these autors have demonstrated that the very internal
regions of ellipticals tend to show an internal isothermal core
quite different from the prediction of a pure de Vaucouleurs law.
Using bright ellipticals in the Virgo cluster they show that the
brightness profile become closer to the de Vaucouleurs law only
after the inner 10 arcsecond region. There is also a prediction of
an external truncation but this could be a more difficult task to
detect since this truncation is heavily dependent of the sky
subtraction procedure.  Finally, we would like to stress that the 
unified treatment for globular clusters and elliptical galaxies 
based on Tsallis distribution as presented here may, in principle,  be extended 
to the framework of the so-called Kaniadakis non-gaussian k-distributions \cite{Kani}.

\section*{Acknowledgments}

JMS is supported by CNPq No. 150484/2007-0. JASL is partially supported by  FAPESP and CNPq (Brazilian Research Agencies) under grants 04/13668-0 and 304792/2003-9, respectively.

\end{document}